\documentclass[]{aa}

\usepackage{soul}
\usepackage{txfonts}
\usepackage{graphicx} 
\usepackage{natbib} 
\usepackage{longtable} 
\usepackage{amssymb}
\usepackage{lscape}
\usepackage{rotating}
\usepackage[english]{babel}
\bibpunct{(}{)}{;}{a}{}{,}

\newcommand{\lppr}{\stackrel{<}{\scriptstyle \sim}}
\newcommand{\lappr}{\raisebox{-0.4ex}{$\lppr$}}

\usepackage[usenames,dvipsnames]{xcolor}
\newcommand{\kms}{${\rm\,km\,s^{-1}}$} 

\begin{document}

\title{The effects of $^{22}$Ne sedimentation and metallicity on the local 40 pc white dwarf luminosity function }

\author{Jordi Tononi\inst{1}\and
        Santiago Torres\inst{1,2}\thanks{Email;
    santiago.torres@upc.edu}\and
        Enrique Garc\'\i a--Berro\inst{1,2}$\dagger$\and
        Mar\'\i a E. Camisassa\inst{3,4}\and 
        Leandro G. Althaus\inst{3,4} \and 
        Alberto Rebassa--Mansergas\inst{1,2}}
        
\institute{Departament de F\'\i sica, 
           Universitat Polit\`ecnica de Catalunya, 
           c/Esteve Terrades 5, 
           08860 Castelldefels, 
           Spain
           \and
           Institute for Space Studies of Catalonia, 
           c/Gran Capita 2--4, 
           Edif. Nexus 104, 
           08034 Barcelona, 
           Spain
           \and
           Facultad de Ciencias Astron\'omicas y Geof\'isicas, 
           Universidad Nacional de La Plata,
           Paseo del Bosque s/n, 
           1900 La Plata, 
           Argentina
           \and
           Instituto de Astrof\'isica de La Plata, UNLP-CONICET,
           Paseo del Bosque s/n, 
           1900 La Plata, 
           Argentina \ \\
           $^{\dagger}$ Deceased 23rd September 2017, 
           }

\date{\today}

\titlerunning{The effects of $^{22}$Ne sedimentation and metallicity}
\authorrunning{Tononi et al.}

\offprints{S. Torres}


\abstract 
{}
%
%
          {We analyze the effect of the sedimentation of $^{22}$Ne on the local white dwarf luminosity function by studying scenarios under different Galactic metallicity models.}
          {We make use of an up-to-date  population synthesis  code based  on Monte  Carlo techniques to derive the synthetic luminosity function. The code incorporates  the most recent and reliable cooling sequences and an accurate modeling of the observational biases, under different scenarios. We first analyzed the case when a constant solar metallicity model is used, and compared the models with and without $^{22}$Ne sedimentation with the observed luminosity function for a pure thin disk population. Then, the possible effects of a thick disk contribution are analyzed. We also studied different metallicity model scenarios including $^{22}$Ne sedimentation.  The analysis is quantified from  an statistical $\chi^2$-test value for the complete, as well as for the  most significant regions of the white dwarf luminosity function. Finally, a best-fit model along with a disk age estimate is derived.}
          {Constant solar metallicity models are not able to simultaneously  reproduce the peak and cut-off of the white dwarf luminosity function. The extra release of energy due to  $^{22}$Ne sedimentation piles up more objects in brighter bins of the faint end of the luminosity function. The contribution of a single burst thick disk population increases the number of stars in the magnitude interval centered around   $M_{\rm bol}=15.75$. Among the metallicity models studied, the one following a Twarog's profile is disposable. Our best fit model was obtained  when a dispersion in metallicities around the solar metallicity value is considered along with a $^{22}$Ne sedimentation model, a thick disk contribution  and an age of the thin disk of $8.8\pm0.2$\,Gyr.}
          { Our population synthesis model is able to reproduce  the local white dwarf luminosity function with a high degree of precision when a dispersion in metallicities around the solar value model is adopted. Although the effects of $^{22}$Ne sedimentation are only marginal and the contribution of a thick disk population is minor, both of them help in better fitting the peak and the cut-off regions of the white dwarf luminosity function.}


\maketitle


\section{Introduction}

White dwarfs are the  most common stellar evolutionary end-point. As a matter of fact, low- and intermediate-mass stars -- namely,  those  with  $M\,\lappr\,8\sim 11\,M_{\sun}$, depending on the metallicity and the intensity of the core overshooting \cite[e.g.][]{Siess2007} --, end their lives as white dwarfs. These stars are very dense objects without relevant nuclear energy sources, where the pressure against gravitational collapse arises essentially from the degenerate electrons. Consequently, white dwarfs are  doomed to a slow and long cooling process. The white dwarf structure is relatively simple and their evolutionary properties are reasonably well understood -- see the review \cite{Althaus2010a} and references therein  for an  in depth  discussion of this  issue. The vast majority of white dwarfs, those of intermediate mass, have a core made of a mixture of C and O plus some impurities,  being  $^{22}$Ne the most abundant. Those white dwarfs with masses $M\la 0.45  \, M_{\sun}$ have He  cores, whilst  those with masses $M \ga 1.1  \, M_{\sun}$ have O-Ne cores. Independently of their inner composition, white dwarf cores are surrounded by a thin non-degenerate envelope having typically $\sim 1 \%$ of  the total mass of the star. Whilst the degenerate core containing the bulk of mass of the white dwarf acts as an energy reservoir, the thin envelope is the responsible for controlling the energy outflow.

In approximately 20\% of white dwarfs  the whole envelope is mainly formed by helium. However, in the remaining $\sim$ 80\% of the cases the helium envelope is  surrounded by an even thinner layer of hydrogen of mass $10^{-4}$ -- $10^{-15}\,  M_{\sun}$. White dwarfs displaying hydrogen lines in their spectra are known  as DA, whilst those characterized by the absence of this feature are generically referred to as non--DA white dwarfs.

As long living and well understood objects from a theoretical point of view, white dwarfs have been used to derive not only important properties of the Galaxy and its components, but also to delimit theoretical cooling models and new physical processes. To cite a few examples, the white dwarf population has been used to study  the  nature  and  history of  the  different components  of our  Galaxy, namely the thin and thick disks  \citep{Winget87, GB88b, GBerro1999, Torres2002, Rowell2011, Rowell2013,Kilic2017}, the  Galactic halo \citep{Mochkovitch1990, Isern1998,GBerro2004,vanOirschot2014,Kilic2019} and more recently the Galactic bulge \citep{Calamida2014,Torres2018}. Moreover, white dwarfs have also been employed in the study of important characteristics, such as the age, subpopulation identification, white dwarf cooling and other relevant parameters, of open and globular clusters -- to cite some of the most representative examples we mention the works of \cite{Salaris2001},  \cite{Calamida2008},  \cite{GBerro2010}, \cite{Jeffery2011}, \cite{Hansen2013}  and \cite{Torres2015}.

A key point in the study of the properties of the white dwarf population is the white dwarf luminosity function --- a comprehensive review can be found in \cite{GBerroOswalt2016}. Initially derived by \cite{Weidemann68}, the white dwarf luminosity function is defined as the number of  white dwarfs  per cubic parsec and bolometric magnitude unit and represents the scenario where  the different ingredients of the white dwarf cooling theory as well as the past history and evolution of the Galaxy  manifest themselves. As previously referenced, the white dwarf luminosity function has been used to determine the age of the Galactic disk and its star formation history as well as to constrain the physics of white dwarf cooling -- including neutrino emission at high temperatures and crystallization processes at relative low core temperatures --, among other examples. Moreover, the white dwarf luminosity function has been also applied to corroborate or discard non-standard physical theories such as the testing of the gravitational constant, $G$, \citep{GBerro1995,GBerro2011}, or as an astro-particle physics laboratory \citep{Isern1992,Isern2008,Dreiner2013,Miller2014}.

The advent of modern large-scale automated surveys and more sophisticated observational techniques has provided us with unprecedented white dwarf samples from which we can test the physics of the white dwarf cooling process. These detailed and more complete available samples are not exclusive to the disk population,  but also to open and globular clusters. In particular, the analysis of the observed white dwarf luminosity function in clusters has provided useful independent determinations of their ages, sometimes in contrast with those derived from main-sequence turn-off stars. As an illustrative example of this issue we can mention the theoretical studies, performed a few decades ago,  of the role of minor chemical species, in particular the sedimentation of the $^{22}$Ne, in the white dwarf cooling    \citep{Isern1991,Isern1997,Isern2000,Bildsten2001}. Unfortunately, the scarcity of complete and statistically significant white dwarf samples at that time, prevented to achieve any conclusive  result and disentangling among different theoretical predictions. It was not until recently that, with the aid of {\sl  Hubble   Space   Telescope} observations, the old metal-rich open cluster NGC\,6791 could be tracked deeply enough for allowing the identification of a clear and significant sample of white dwarfs from which it was possible to obtain its luminosity function \citep{Bedin2005,Bedin2008}. The initial discrepancy between the NGC\,6791 age determined by the main-sequence turn-off point ($\sim 8\,$Gyr) and the age derived from the termination of the white dwarf cooling sequence ($\sim 6\,$Gyr) could only be reconciled once the sedimentation of $^{22}$Ne along with detailed models of phase separation during the crystallization of typical CO white dwarfs were taken into account \citep{GBerro2010}. 

The main physical reason underneath of the effects induced by $^{22}$Ne in the white dwarf cooling rises from the neutron excess of this isotope. $^{22}$Ne nuclei are formed as a result of helium captures on $^{14}$N left from hydrogen burning via the CNO cycle, through the reactions $^{14}{\rm N}(\alpha, \gamma)^{18}{\rm F}(\beta^+)^{18}{\rm O}(\alpha, \gamma)^{22}$Ne.  The excess of neutrons (two relative to the predominant $A = 2Z$ nuclei) of  $^{22}$Ne causes a downward force on them, biasing its diffusive equilibrium and  sinking these nuclei into the deep interior of the white dwarfs \citep[e.g][]{Bildsten2001}. This fact is  responsible for rapid sedimentation of $^{22}$Ne in the interior of white dwarfs \citep{Isern1991,Deloye2002,Althaus2010,Camisassa2016}. This diffusion process releases energy, yielding a marked delay in the cooling times of white dwarf in solar and super solar metallicity environments. Particularly, in white dwarfs arising from solar metallicity progenitors, the delays in cooling times reach about 1 Gyr, (see e.g. \citep{Camisassa2016}. These delays are expected to alter the local white dwarf luminosity function.  The impact of $^{22}$Ne diffusion is usually ignored in the evolutionary calculations existing in the literature, and thus, not considered in the population synthesis studies performed on the local sample of white dwarfs. In this study, we aim to
asses the effects of this process on the white dwarf luminosity function of the 40 pc local sample and to estimate the accuracy of the predictions of the usual calculations which ignore this process.

Our  paper is  organized as  follows.  In  Sect.~\ref{sec:obsdat} we
describe in detail  the  set  of  observations to  which  our theoretical  simulations will be  compared. A summarized description of our population synthesis code along with its main physical inputs is provided in Sect.~\ref{sec:popsyn}. In Sect.~\ref{sec:metmod} we describe the different metallicity models considered in this study and in Sect.~\ref{sec:res} we present the results of our Monte Carlo population  synthesis calculations. In  particular, in Sect.~\ref{sec:e22ne}  we compare  the models considering and excluding $^{22}$Ne sedimentation.  Sect.~\ref{sec:refmod} is  devoted  to analyze the effects of $^{22}$Ne sedimentation when  a metallicity dispersion is considered. In Sect.~\ref{sec:casa} we analyze the observational metallicity models and in Sect.~\ref{sec:bestfit} we present our best fit model. Finally, Sect.~\ref{sec:con} summarizes our main results and discusses their significance.

\section{The observational sample}
\label{sec:obsdat}

As previously stated, the white dwarf luminosity function represents a capital tool in the study of white dwarf evolutionary physical processes as well as it provides an inexhaustible source of information of the evolutionary history of our Galaxy.  Obviously, a key point in these studies is the comparison between the theoretically derived and the observed luminosity functions. To that end, a statistically significant and complete luminosity function is essential for analyzing the properties of the white dwarf luminosity function. 

Recent magnitude- and proper motion-limited samples, like the SDSS \citep{Harris2006} or the SuperCOSMOS Sky Survey \citep{Rowell2011,Rowell2013} respectively, have substantially increased the number of known white dwarfs over the last decade. This has allowed going deeper into the Galaxy up to several hundred parsecs. However, these samples are severely affected by observational biases, completeness issues and selection procedures, which must be necessarily taken into account in any detailed analysis. Moreover, the detection of faint objects is inherently difficult in magnitude-limited surveys and, as a consequence, these samples suffer from a paucity of white dwarfs at the faintest bins of the luminosity function. This problem is especially important since valuable information is enclosed on those faint bins. Even with the unquestionable improvement in data that {\sl Gaia} has provided (e.g. \cite{Hollands2018,Jimenez2018}, the resulting white dwarf sample will be no exempt from such biases -- the limiting magnitude, the parallax errors, the photometric flux excess factors, and other criteria applied in selecting Gaia white dwarf samples can introduce a source of incompleteness for the faintest luminosity bins    \citep[e.g.][]{Torres2005,Barstow2014,Jimenez2018}--,  besides the fact that a throughly spectroscopic study is needed in order to obtain accurate luminosity estimates.

On the other hand, volume-limited samples are the best approach to an  effectively complete sample. In this sense, \cite{Holberg2008} and \cite{Giammichele2012} studied the white dwarf population within 20 pc of the Sun for an unbiased and nearly complete sample of $\sim$130 white dwarfs. More recently, \cite{Holberg2016} extended the survey to 25 pc. This last sample included 232 objects, however the global estimate for its completeness was no greater than a $70\%$.

Nevertheless, our choice for this study was the white dwarf population within 40 pc of the Sun from \cite{Limoges2015}. This compilation of white dwarfs, extracted from the SUPERBLINK survey, corresponds to that of a magnitude- and proper-motion limited sample, which contains $\sim 500$ northern hemisphere objects. Although the sample is magnitude-limited, it is expected to be $\sim70\%$ complete. In fact, a detailed analysis of selection criteria effects indicates that the average completeness for objects  brighter than $M_{\rm bol}=14$ is close to $85\%$, while there is a strong deficit of objects for magnitudes larger than $M_{\rm bol}>16$ \citep{Torres2016}. In spite of this, it is possible to resolve the observed drop-off of the luminosity function as an unambiguous consequence of the finite age of the Galactic disk. Hence, the 40 pc sample of \cite{Limoges2015} both gathers an acceptable degree of completeness and an statistically significant number of objects --- larger than current volume-limited surveys. For instance, the 40 pc \cite{Limoges2015} sample contains 492 objects, while the practically complete 20 pc sample of \cite{Hollands2018} only 130 stars. This permits resolving many of the main characteristics of the Galactic history through its luminosity function and, at the same time it is completely suitable for studying in detail the evolutionary cooling of white dwarfs.

\section{The population synthesis code}
\label{sec:popsyn}

A detailed description  of the main ingredients employed  in our Monte Carlo population  synthesis code  can be found  in our  previous works \citep{GBerro1999,Torres2001, Torres2002,GBerro2004}. Here, we briefly describe the most important characteristics of our simulator in the modeling of the thin and thick disk populations. In addition, we provide details about the evolutionary sequences used in this work. 
 
First of all, we spatially distributed our stars randomly generating their positions in a spherical region centered on the Sun and adopting a radius of $50$~pc. We used a double exponential distribution for the  local density of stars. The spatial distribution perpendicular to the Galactic plane followed an exponential profile with an adopted constant Galactic scale height of 250~pc for the thin disk population and  1.5\,kpc for the thick disk. Similarly the distribution in the Galactic plane was generated according to a constant  scale length  of  2.6~kpc and 3.5~kpc for the thin and thick disk, respectively. Secondly, the time at which each synthetic  star was born was generated according to a constant star formation rate once an age of the Galactic thin disk, $t_{\rm disk}$, is adopted.  A burst of star formation happened 0.6\,Gyr ago is also introduced in order to reproduce the excess of hot objects \citep{Torres2016}. The thick disk is modeled according to a single burst of star formation \citep[e.g.][]{Reid2005} for which we adopt a 1\,Gyr of burst duration. The age of the thick disk is assumed to be 1.6\,Gyr older than the thin disk (see \cite{Kilic2017} and Section \ref{s:thick} for a further discussion). In parallel, the  mass of each star  was drawn according to  a Salpeter mass function  \citep{Salpeter1955} with an exponent $\alpha=-2.35$. This prescription, for the relevant range of masses studied here, is equivalent to the standard initial mass function of \cite{Kroupa_2001}. Once our synthetic thin disk star is formed we associated to it a metallicity according to a certain metallicity law (see Section \ref{sec:metmod}).  In all cases, a  metallicity of $[{\rm Fe/H}]\approx -0.7\,$(dex) is adopted for thick disk stars.  The evolutionary ages  of the progenitors were those of \cite{PrivAlthaus} which, for the range of masses and metallicities used here, are equivalent to those of BaSTI\footnote{http://basti.oa-teramo.inaf.it/} models. Knowing the  age of the  Galactic disk and the age, metallicity, and mass of the progenitor stars, we know which of those stars had time to become white dwarfs. In these cases, we self-consistently derived the white dwarf masses from the evolutionary tracks of \cite{Renedo10}, which are equivalent to use the semi-empirical initial-to-final mass relation of \cite{cat2008}.  We also randomly assigned an atmospheric composition to each artificial white dwarf. In particular, we adopted the canonical fraction of 80\% of white dwarfs with pure hydrogen atmospheres, and assumed the remaining objects to be pure helium atmospheres. Finally, velocities for each star were randomly chosen taking into account the differential rotation of the Galaxy and the peculiar velocity of the Sun, $(U_{\sun},\,V_{\sun},\,W_{\sun})=(7.90,\,11.73,\,7.39)\,$\kms \citep{Bobylev2017}. Mean Galactic velocity values respect to the LSR and
their dispersions for the thin and thick disk populations are those from \cite{Torres2019} -- Table 2.

The set of adopted cooling sequences employed here encompasses the most recent evolutionary calculations for different white dwarf masses. For white dwarfs  masses smaller than  $1.1\, M_{\sun}$ we adopted the cooling tracks of H-rich atmosphere and carbon-oxygen cores of \cite{Camisassa2016}, suitable for solar metallicity populations. These recent cooling tracks are the result of the full evolutionary calculations of their progenitor stars, starting at the Zero Age Main Sequence (ZAMS), all the way through central hydrogen and helium burning, thermally-pulsing AGB and post-AGB phases. For H-deficient white dwarfs with masses smaller  than  $1.1\, M_{\sun}$ we calculated additionally new cooling sequences for the purpose of this work, evolved from the ZAMS through the born-again scenario  for solar metallicity progenitors, as described in \cite{Camisassa2017}. Subsolar metallicity DA white dwarf cooling tracks were those of \cite{Althaus2015}  which consider residual hydrogen shell burning. In all cases, the cooling tracks take into account the energy released by latent heat and include the separation phase of carbon and oxygen due to crystallization, following the phase diagram of \cite{Horowitz2010}. Additionally, the solar metallicity cooling tracks  (for both H-rich and H-deficient atmospheres) above described also take into account the sedimentation of $^{22}$Ne nuclei, employing the new diffusion coefficients based on molecular dynamics simulations of \cite{Hughto2010}. For white dwarf masses larger  than $1.1\, M_{\sun}$  we used  the evolutionary sequences for DA oxygen-neon  white   dwarfs  of  \cite{Alt2005}   and  \cite{Alt2007}. Finally, for each white dwarf we interpolated the  luminosity, effective  temperature, and the value of  $\log g$, together with all the relevant parameters, in the corresponding white dwarf evolutionary track.  We also interpolated their $UBVRI$  colors, which  we then converted  to the  $ugriz$ color system  in order to apply the selection criteria of the observed sample \citep{Limoges2015,Torres2016}.

The final synthetic white dwarf population for each of our models, is the resulting of $50$ independent Monte Carlo simulations of different initial seeds  and normalized to the exact number (492 objects) of the observed sample (see Sect.\,\ref{sec:obsdat}). Each particular simulation contained a number of synthetic white dwarfs around that number of objects. This way, convergence in all the final values of the relevant quantities can be ensured.

\section{Metallicity models}
\label{sec:metmod}

Metallicity is a fundamental parameter that not only rules the evolutionary lifetime of stars, but it is also deeply linked to the Galactic evolutionary history. Consequently, it is important to understand the influence of metallicity in the white dwarf luminosity function and, in particular, in the estimation of the Galactic disk's age. A recent analysis by \cite{Cojocaru2014}  shows that, regardless of the choice of the metallicity law, the estimation of the disk age derived from the white dwarf luminosity function is a robust method given that the cut-off remains unchanged. Moreover, this study  shows that neither the shape of the bright portion of the white dwarf luminosity function nor the position of its cut-off at low luminosities is affected by the assumed metallicity law or the ratio of DA to non-DA white dwarfs. It is also worth mentioning that \cite{Rebassa2016} recently analyzed a pilot sample of 23 white dwarfs in binary systems with main sequence companions. From these objects they derived accurate white dwarf ages and main sequence star metallicities and found that there is not a clear correlation between age and metallicity at young and intermediate ages (0-7 Gyr). This large scatter of metallicity values is also extensible for older ages as observed when analysing single main sequence stars samples \citep{Casagrande2011,Casagrande2016,Haywood2013,Bergemann2014}. This implies that some physical  mechanism is underneath the observed scatter of the age-metallicity relation.

Despite that at first approximation metallicity seems not to play an important role in the white dwarf luminosity function, the observed scatter of metallicity values may induce some variations in the cooling evolution of individual stars. This is precisely the effect that we aim to study here,  in view of the fact that detailed white dwarf evolutionary sequences, as previously shown, suggest appreciably changes in the cooling rate as a function of the metallicity. It is also important to note that a smaller sample may be more affected by local inhomogeneities (e.g. moving groups, stars associations) and thus the effects of a scatter in metallicities could be enlarged. In contrast, metallicity values are averaged on larger magnitude-limited samples \cite[e.g.][]{Cojocaru2014} but, nevertheless, the completeness of these samples are smaller. For all these reasons we consider important to extend our study of the local sample of white dwarfs to other metallicity models than those used by \cite{Cojocaru2014}.  

The models used in the present study are as follows. Our first model assumes a constant solar metallicity of $Z_{\sun}=0.014$ independently of the age of the star. This will be our reference model for the rest of the analysis. Our second model considers a dispersion along the solar metallicity value of $\sigma[{\rm Fe/H}]\approx 0.4$(dex), which implies that the  range of $Z$ is expanded from  subsolar values of $\approx 0.003$ up to  supersolar values of $\approx 0.05$. This dispersion is in agreement with the  data collected by Geneva-Copenhagen survey for isolated main sequence stars \citep{Casagrande2011} as well as with more recent seismic ages, obtained using red giants observed by Kepler \citep{Casagrande2016}. It is also in complete agreement with the age-metallicity relation specifically derived for white dwarfs by \cite{Rebassa2016}.

The third model is based on the observed age-metallicity relation provided by \citet{Casagrande2011,Casagrande2016,Haywood2013,Bergemann2014}. In particular, we consider a gradual decrease of [Fe/H] for objects older than 8 Gyrs. Consequently, this model considers that the first stars in the Galaxy have $[{\rm Fe/H}]\approx -0.3\,$(dex), and that [Fe/H] linearly increases until achieving a solar metallicity value with a dispersion of $\sigma[{\rm Fe/H}]\approx 0.35\,$(dex). For stars with ages below 8 Gyr, this third model considers [Fe/H] to remain constant and equal to the solar metallicity with a dispersion of $\sigma[{\rm Fe/H}]\approx 0.4\,$(dex). Finally, the fourth and last model assumes the classic \cite{Twarog1980}'s age-metallicity relation. That is, a law that predicts a monotonous increase of [Fe/H] that begins with a zero value for the oldest stars and that finishes with a solar metallicity value for present day stars. A dispersion of $\sigma[{\rm Fe/H}]\approx 0.1\,$(dex) is added to the mean value at a given age.

\section{Results}
\label{sec:res}

\begin{figure}
   {\includegraphics[width=1.05\columnwidth]{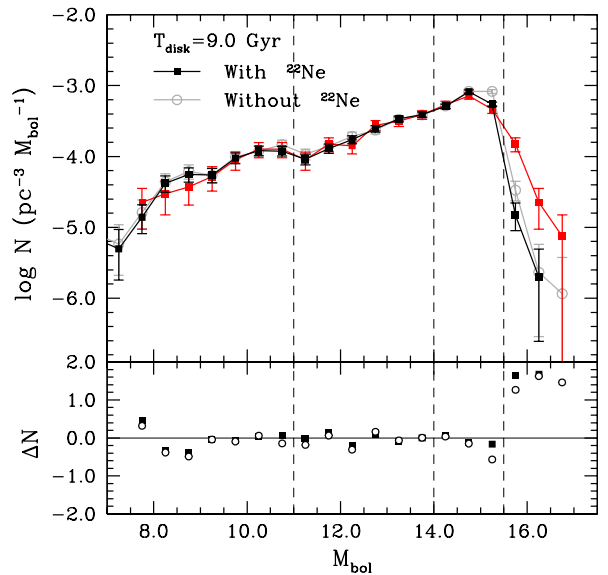}}
\caption{Synthetic white dwarf luminosity functions that result from the models including and disregarding $^{22}$Ne sedimentation (black solid squares and gray  open circles, respectively) and assuming a constant solar-metallicity model, compared with the observed white dwarf luminosity function (red lines) of \cite{Limoges2015}. In the bottom panel we show the residuals between the simulated and the observed samples. See text for details.}
\label{f:figura1}
\end{figure}

\subsection{The effects of $^{22}$Ne sedimentation in the thin disk population}
\label{sec:e22ne}

First of all we analyze the effects of $^{22}$Ne sedimentation just comparing  the models that include and disregard this feature  for a pure thin disk population. In both cases we  fix the metallicity to the constant standard solar value and adopt a disk age of  $T_{\rm disk}=9.0\,$Gyr in accordance with \cite{Torres2016}. It is worth noting that 
this age value represents an indicative guess. There exists a wide spread in ages for the 40 pc sample of \cite{Limoges2015} depending on the model assumptions (for instance, the main-sequence lifetime or the initial-to-final mass relation adopted) or the estimate contamination in the cut-off region from the thick disk population. Lower limits for a thin disk population  can be as short as $6.8\,$Gyr \citep{Kilic2017}, while upper limits find a disk age of around $11\,$Gyr \citep{Limoges2015}. Anyhow, for our purpose here it is enough to adopt a reasonable guess and we postpone to the next Sections a more detail analysis of the effects of the age and the thick disk population.

The results are shown in Figure \ref{f:figura1} where we plotted the observed white dwarf luminosity function (solid squares and red lines) of \cite{Limoges2015} compared with the model that incorporates the $^{22}$Ne sedimentation (solid squares and black lines) and the one without the $^{22}$Ne sedimentation (open circles and gray lines). Additionally, in the bottom panel of Fig. \ref{f:figura1} we display the residuals \citep[e.g.][]{Cojocaru2014} for both the observed and simulated luminosity functions, which better helps in assessing the differences between both simulation, defined as
\begin{equation}
\Delta N=2\frac{N_{\rm obs}-N_{\rm sim}}{N_{\rm obs}+N_{\rm sim}}
\end{equation}
where $N_{\rm obs}$ stands for the number of objects per bin of the observed sample and $N_{\rm sim}$ for the corresponding synthetic simulated sample.

Moreover, in order to find the best fit in a quantitative way for our models and to help in the subsequent discussion, we performed a multiple $\chi^2$ test. Recalling that in the white dwarf luminosity function there exists a difference of nearly two orders of magnitude in the number of objects between the peak and the less populated bins, it is clear that a simple statistical test will underestimate the physical characteristics of the less populated regions. For this reason, we considered not only the entire aspect of the luminosity function, but also we divided it into four regions of relevant physical importance, namely the hot branch ($M_{\rm bol}\leq 11.0$) related to a possible recent burst of star formation, a middle region with  constant slope ($11.0 <M_{\rm bol}\leq 14.0$), the peak region ($14.0<M_{\rm bol}\leq 15.5$) and the cut-off region ($M_{\rm bol}> 15.5$) which  contains valuable information on the initial formation of the disk. These regions, that we will call burst, middle, peak and cut-off are delimited by a dashed vertical line and easily identified from left to right, respectively, in Fig. \ref{f:figura1}. Once obtained the $\chi^2_{\rm i}$ for each of these regions, we derive the reduced value as  $\chi^2_{\rm r,i}\equiv\chi^2_{\rm i}/\nu$, where $\nu$ is the  degree of freedom defined as the number of observations minus the number of constrains. In our case, the number of constrains is one given that the synthetic samples are normalized to the same number of objects than the observed one.  Thus, we get  $\nu=6$, 5, 2, 2, and 18, for the burst, middle, peak, cut-off and entire luminosity function, respectively. Additionally we use our Monte Carlo simulator to estimate the error deviation per bin. This fact allow us to calibrate our $\chi^2_{\rm r,i}$, hence  avoiding exceptionally smaller values of $\chi^2_{\rm r,i}$ due to an overestimation of the errors.  It is also worth saying that the $\chi^2$-test performed here shall not be interpreted as an absolute measure of the adjustment, rather as a tool for comparative purposes among models. For further discussion of the $\chi^2$ capabilities and other statistical methods we point the reader to \citet{Andrae2010,Feigelson2012}.

\begin{figure}
   \resizebox{\hsize}{!}
   {\includegraphics[width=\columnwidth]{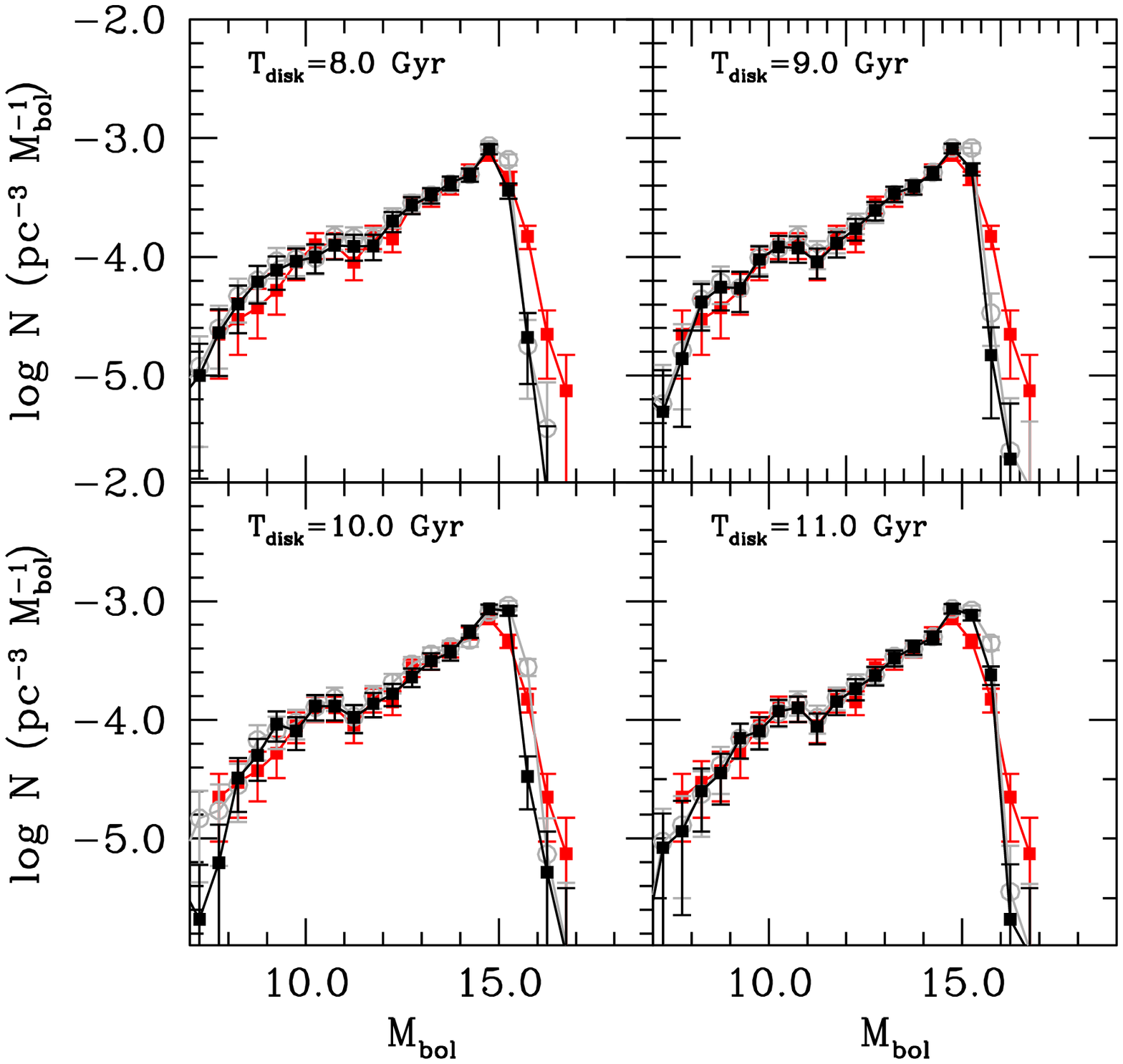}}
\caption{Synthetic white dwarf luminosity function (black lines) for different disk ages and a constant solar-metallicity and a pure thin disk} model, compared with the observed 
white dwarf luminosity function (red lines) of \cite{Limoges2015}.
\label{f:model1}
\end{figure}

At first glance, Fig.\ref{f:figura1} highlights a reasonable good global agreement between the simulated and the observed samples, despite the fact that some discrepancies arise in the peak region for the model without $^{22}$Ne diffusion and in the cut-off region for both models. The slope of the luminosity function for hot and moderate luminosities is suitably reproduced by both synthetic functions as derived by the close to 1 values of the reduced $\chi^2$-test in these regions ($\chi^2_{\rm r,burst}=1.19$ and $\chi^2_{\rm r,middle}=1.16$ for the model with  $^{22}$Ne diffusion, and $\chi^2_{\rm r,burst}=1.31$ and $\chi^2_{\rm r,middle}=1.40$ for the model without). This fact is borne out by the almost
zero values  of the residual for the $^{22}$Ne sedimentation model (solid black squares) and the model without this effect (open black circles), see bottom panel of Fig. \ref{f:figura1}. It is only in the peak and in the cut-off region of the luminosity function where discrepancies arise. In the peak region, the residuals indicate the sample resulting from the model that incorporates $^{22}$Ne sedimentation agrees substantially better with the observed one. The reduced $\chi^2$-test in this region, $\chi^2_{\rm r,peak}=1.10$ and $\chi^2_{\rm r,peak}=3.24$ for the models with and without  $^{22}$Ne diffusion, respectively, also confirm that result. A visual inspection of the peak region from Fig. \ref{f:figura1} clearly shows that the model with $^{22}$Ne sedimentation resembles notably well the observed shape, while the model that ignored $^{22}$Ne diffusion continues with an increasing slope, even far beyond the observed maximum. This fact is a direct consequence of the extra release of energy due to the diffusion of  $^{22}$Ne, which induces a delay in the cooling times of white dwarfs. Consequently, for a fixed disk age, more objects remain in brighter luminosity bins when diffusion of  $^{22}$Ne is taken into account, while on the contrary, objects reach lower luminosities when $^{22}$Ne diffusion is ignored. Concerning the cut-off region, at least for a disk age of 9.0\,Gyr and for a pure thin disk population, both synthetic models seem far away to reproduce the observed values ($\chi^2_{\rm r,cut-off}=13.87$ and $\chi^2_{\rm r,cut-off}=9,82$, for the models with and without $^{22}$Ne diffusion, respectively). Finally, when the entire luminosity function is considered, the model with  $^{22}$Ne sedimentation better reproduces the observed luminosity function ( $\chi^2_{\rm r,entire}=3.04$ for the model with, and $\chi^2_{\rm r,entire}=4.09$ for the model without $^{22}$Ne sedimentation). However, both $\chi^2$ test values are far from being considered an adequate adjustment.

Previously to study the effect of the contamination of the thick disk population, we analyze in more detail the effect of the age of the thin disk population in our models with and without  $^{22}$Ne sedimentation. We derive the corresponding luminosity function for the range of ages between 7\,and 12\,Gyr, while in Figure \ref{f:model1} we plot the corresponding luminosity functions for those ages between 8\, and 11\,Gyr. The complete set of values for the $\chi^2$-test of the different regions, model and ages is presented in Table \ref{tab:model1}.  For comparative purposes, we also show the observational luminosity function (red lines) of \cite{Limoges2015}. The results obtained reinforce our previous idea that neither of the two models is able to reproduce the cut-off and the peak simultaneously. The best fit of the cut-off is achieved for an age of 11\,Gyr for the model with  $^{22}$Ne diffusion and for an age of 10\,Gyr for the model without, being this fact a clear consequence of the extra release of energy of the   $^{22}$Ne sedimentation models. However, at these ages the peak region is inadequately resolved in both models.  On the other hand, the peak region is well resolved for an age of 9\,Gyr when $^{22}$Ne difussion is taken into account, while models that ignored it need ages $<7\,$Gyr to obtain a reasonable fitting.

Summing up, the effects  of the  $^{22}$Ne sedimentation are negligible  in the hot  and medium region of the white dwarf luminosity function, while  neither of the two models, with or without  $^{22}$Ne sedimentation, are able to reproduce the observed low-luminosity bins when a pure thin disk model and constant metallicity are considered. However, we found hints that the model which include  $^{22}$Ne sedimentation is, in principle, more plausible to reproduce the observed peak region.

\begin{table*}
\caption{Reduced $\chi^2$ test values for the different regions as well as for the entire luminosity function and for the different parameters of our models. See text for details.}
\smallskip
\begin{center}
{\small
\begin{tabular}{ccccccccc}  
\hline
\noalign{\smallskip}
$T_{\rm disk}$\,(Gyr) & $^{22}$Ne diff. & Metalliciy & Population &  $\chi^{2}_{\rm r,burst}$  & $\chi^{2}_{\rm r,middle}$ & $\chi^{2}_{\rm r,peak}$ & $\chi^{2}_{\rm r,cut-off}$ & $\chi^{2}_{\rm r,entire}$ \\
\noalign{\smallskip}
\hline
\noalign{\smallskip}
7 & yes & Model 1 & pure thin & 1.33 &	1.54 &	7.15 & 14.26 & 7.62 \\
8 & yes & Model 1  & pure thin & 1.43 &	1.61 & 1.22 & 12.78 & 3.27 \\
9 & yes & Model 1  & pure thin & 1.19 & 1.16 & 1.10 & 13.87 & 3.04  \\
10 & yes & Model 1  & pure thin & 1.28 & 1.20 &	4.02 & 11.02 & 4.73  \\
11 & yes & Model 1  & pure thin & 1.17 & 1.14 & 2.76 & 4.62 & 2.81 \\
12 & yes & Model 1  & pure thin & 1.03 & 1.40 & 3.13 & 11.31 & 4.10  \\
\noalign{\smallskip}
\hline
\noalign{\smallskip}
7 & no & Model 1  & pure thin & 1.73 & 1.94 & 1.63 & 13.78 & 4.93 \\
8 & no & Model 1  & pure thin & 1.78 & 2.00 & 1.88 & 12.84 & 4.08  \\
9 & no & Model 1 & pure thin & 1.31 & 1.40 & 3.24 & 9.82 & 4.09  \\
10 & no & Model 1  & pure thin & 1.51 & 1.54 & 4.35 & 5.53 & 4.37  \\
11 & no & Model 1  & pure thin & 1.18 & 1.47 & 3.47 & 17.00 & 5.33 \\
12 & no & Model 1  & pure thin & 1.00 &	1.56 & 3.96 & 21.59 & 6.34  \\
 \noalign{\smallskip}
\hline
\noalign{\smallskip}
7 & yes & Model 1 & thin + thick & 1.17 & 1.35 & 4.40 & 15.39 & 5.68  \\
8 & yes & Model 1  & thin + thick & 1.11 & 1.23 & 0.95 & 10.09 & 2.34  \\
9 & yes & Model 1  & thin + thick & 0.99 & 1.21 & 1.70 & 6.19 & 2.20  \\
10 & yes & Model 1  & thin + thick & 1.14 & 1.36 & 3.81 & 2.99 & 3.33  \\
11 & yes & Model 1  &thin + thick & 1.04 & 1.53 & 4.29 & 4.51 & 3.92  \\
12 & yes & Model 1  & thin + thick & 1.51 & 1.42 & 3.67 & 11.09 & 4.70  \\
\noalign{\smallskip}
\hline
\noalign{\smallskip}
7 & no & Model 1  & thin + thick & 1.22 & 1.38 & 1.02 & 14.36 & 3.18

\\
8 & no & Model 1  & thin + thick & 1.16 & 1.34 & 2.66 & 9.76 & 3.57  \\
9 & no & Model 1  & thin + thick & 1.08 & 1.18 & 5.02 & 5.20 & 4.42 \\
10 & no & Model 1  & thin + thick & 1.16 & 1.59 & 4.34 & 11.31 & 5.10  \\
11 & no & Model 1  & thin + thick & 1.16 & 1.80 & 5.55 & 23.29 & 7.90   \\
12 & no & Model 1  & thin + thick & 1.71 & 2.09 & 4.41 & 24.77 & 7.73 \\

\noalign{\smallskip}
\hline\
\end{tabular}
}
\label{tab:model1}
\end{center}
\end{table*}

\subsection{The effects of a thick disk population}
\label{s:thick}

It is expected an approximately $20\%$ contribution of the thick disk population to the mass budged of white dwarfs in the solar neighbourhood \citep[e.g][]{Reid2005}. Moreover, in a recent population classification study, \cite{Torres2019} estimate that for the nearly complete {\it Gaia}-DR2 white dwarf sample within 100\,pc, the contribution of the thick disk population is  $25\%$, increasing up to $35\%$ for the faint magnitude regime. Consequently, the inclusion of the thick disk population is expected to be relevant for a good modeling of the cut-off of the luminosity function. As explained in Section  \ref{sec:popsyn}, we introduce a low-metallicity old-single burst population, adopting a fraction of  $25\%$ in the number of thick disk stars. For the age of the thick disk population we follow the result argued by \cite{Kilic2017}, i.e. the age of the thick disk is 1.6\,Gyr older than the thin disk age. This permits us to use only one free parameter, i.e. the thin disk age, as long as the difference between the two Galactic components remains constant. 

\begin{figure}
   \resizebox{\hsize}{!}
   {\includegraphics[width=\columnwidth]{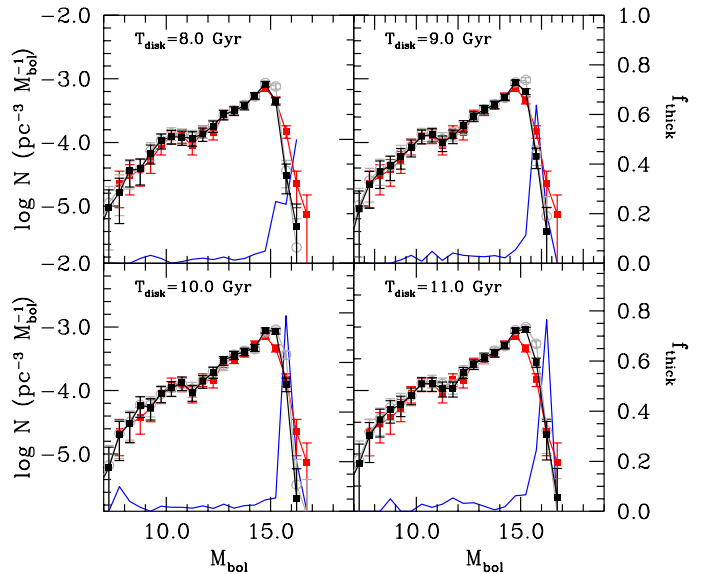}}
\caption{Same as Fig.\ref{f:model1} but considering a thin plus thick disk population. Also plotted (blue continuous line) the fraction of thick disk white dwarfs (right axis). See text for details. }
\label{f:flthick}
\end{figure}

The results  obtained for the thin plus thick disk population with and without $^{22}$Ne sedimentation  along with their corresponding reduced $\chi^2$-test values for the different regions and the entire luminosity function are presented in Figures \ref{f:flthick} and Table \ref{tab:model1}. Additionally, we plot (blued continues line and right axis) the fraction of thick disk white dwarfs, $f_{\rm thick}$, defined as the number of thick disk white dwarfs with respect to the total number of objects for the same magnitude bins as the luminosity function. A comparative look of  Figures \ref{f:model1} and \ref{f:flthick} reveals only marginal changes in the white dwarf luminosity function when the thick disk population is taken into account. In particular, the burst, middle and peak regions of the luminosity function are not affected by the thick disk population. As expected, only in the cut-off region are observable some slight changes, but only for a given interval of bolometric magnitude. For instance, for a disk age of $9\,$Gyr, the bin around $M_{\rm bol}=16.0$ is dominated by thick disk white dwarfs ($f_{\rm thick}>60\,\%$). However, other bins of the cut-off region remain unaffected. This fact is a consequence of the adopted single burst formation model for the thick disk population. As it is well known, single bursts of star formation manifest as peaks in the luminosity function \citep{NohScalo1990}. Consequently, the thick disk population, modeled as an old single burst of formation, produces an increment of these objects only for a certain magnitude interval. Even though we have modelled the burst along $1\,$Gyr of star formation, this is not enough for extending the thick disk contribution to more than a pair of bins in the luminosity function. On the other hand, the larger the burst duration the less intense is the corresponding peak in the luminosity function, that is, smaller values of the fraction $f_{\rm thick}$. 

Nevertheless, although the inclusion of the thick disk population is not able to adequately reproduce the cut-off of the observed white dwarf luminosity function, it seems to help in the overall fitting. In fact, the best fit for the models analyzed so  far (see Table \ref{tab:model1}) corresponds to a model that includes the thick disk population, takes into account $^{22}$Ne diffusion and for a disk age of $9\,$Gyr ($\chi^2_{\rm entire}=2.20$). Although a thorough analysis of the thick disk contribution is beyond the scope of the present work, we have checked other possibilities. For instance,  \citep{Kilic2017} derived best-fit ages of 6.8 Gyr for the thin disk and 8.7 Gyr for the thick disk. Adopting these ages,  our population synthesis analysis derives a reduced $\chi^2$ value for the entire luminosity function of $\chi^2_{\rm r,entire}=5.91$ and $\chi^2_{\rm r,entire}=2.92$ for the cases with and without $^{22}$Ne diffusion, respectively. However, for ages of 8.2 and 10.1 Gyr for the thin and thick disk, respectively, the model with $^{22}$Ne diffusion achieves a better fit with $\chi^2_{\rm r,entire}=2.28$, while the case without is $\chi^2_{\rm r,entire}=3.12$. Other possibilities include fitting separately the ages of the thin and thick disk from the peak and the cut-off, respectively, and leaving the fraction of thick disk stars as a free parameter. However, for a reasonable range of ages and values of the thick disk contribution, the final fitting is not substantially improved with respect to our best fit. Consequently, in what follows we study the effect of the metallicity for those models that include  $^{22}$Ne diffusion and a thick disk population as adopted in the best-fit model obtained so far.

\begin{figure}
   \resizebox{\hsize}{!}
   {\includegraphics[width=\columnwidth]{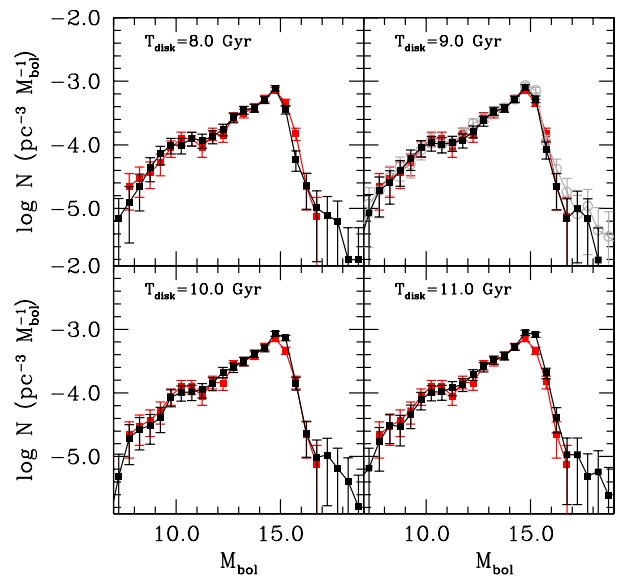}}
\caption{Same as Fig.\ref{f:model1} but considering a thin plus thick population and metallicity model with a dispersion around the solar metallicity value (Model 2). See text for details.}
\label{f:model2}
\end{figure}

\subsection{The effects of a metallicity dispersion}
\label{sec:refmod}

We continue by analyzing  the effects of adding a constant metallicity dispersion around the solar-metallicity value (Model 2).  The derived luminosity functions for different disk ages are shown in Figure \ref{f:model2} (black solid lines), respectively,  while for comparative purposes the observational luminosity function (red lines) of \cite{Limoges2015} in also shown. 

The first evident trend that we observe for Model 2 is that the extended tail of the synthetic luminosity function (Figure \ref{f:model2}) perfectly matches  the drop-off region of the observed luminosity function, in particular for a disk age of $9\,$Gyr. A series of factors seem to contribute to this fact. First of all, larger metallicity values reduce the lifetime of white dwarf progenitors thus promoting more objects to the faint end of the luminosity function, especially those more massive. Secondly,  the extra release of energy provided by $^{22}$Ne sedimentation is more intense for supersolar metallicities. Consequently, more objects are piled up in brighter bins when  $^{22}$Ne sedimentation is taken into account. This last fact can be visualized in Fig. \ref{f:model2} where in panel for disk age $9\,$Gyr we have superimposed the luminosity function (gray line open symbols) for the case when $^{22}$Ne diffusion is omitted. Even though the change is minor, the inclusion of $^{22}$Ne sedimentation in the models permits to slight shift objects to brighter bins and then perfectly match the observed distribution.

\begin{table*}
\caption{Same as Table \ref{tab:model1} but for different metallicity models.}
\smallskip
\begin{center}
{\small
\begin{tabular}{ccccccccc}  
\hline
\noalign{\smallskip}
$T_{\rm disk}$\,(Gyr) & $^{22}$Ne diff. & Metalliciy & Population &  $\chi^{2}_{\rm r,burst}$  & $\chi^{2}_{\rm r,middle}$ & $\chi^{2}_{\rm r,peak}$ & $\chi^{2}_{\rm r,cut-off}$ & $\chi^{2}_{\rm r,entire}$ \\
\noalign{\smallskip}
\hline
\noalign{\smallskip}
7 & yes & Model 2 &  thin + thick & 1.15 & 1.39 & 3.04 & 7.27 & 3.49  \\
8 & yes & Model 2  & thin + thick & 1.30 & 1.29 & 1.13 & 4.27 & 1.69 \\
9 & yes & Model 2  &  thin + thick & 1.11 & 1.28 & 0.95 & 2.11 & 1.12  \\
10 & yes & Model 2  &  thin + thick & 1.17 & 1.60 & 2.40 & 1.00 & 2.04  \\
11 & yes & Model 2  &  thin + thick & 1.18 & 1.43 & 3.41 & 1.88 & 2.92  \\
12 & yes & Model 2  &  thin + thick& 1.11 & 1.44 & 3.26 & 8.08 & 3.75   \\
\hline
\noalign{\smallskip}
7 & yes & Model 3 &  thin + thick& 	1.15 & 1.39 & 3.04 & 7.38 & 3.49  \\
8 & yes & Model 3  &  thin + thick & 1.30 & 1.29 & 1.13 & 4.27 & 1.69  \\
9 & yes & Model 3  &  thin + thick & 1.10 & 1.32 & 1.24 & 1.89 & 1.30  \\
10 & yes & Model 3  &  thin + thick & 1.18 & 1.20 & 2.17 & 1.00 & 1.78  \\
11 & yes & Model 3  &  thin + thick & 1.14 & 1.30 & 2.24 & 2.25 & 2.08  \\
12 & yes & Model 3  &  thin + thick & 1.14 & 1.53 & 4.46 & 10.49 & 5.03  \\
\noalign{\smallskip}
\hline
\noalign{\smallskip}
7 & yes & Model 4 &  thin + thick & 1.05 & 1.39 & 5.06 & 13.30 & 5.78  \\
8 & yes & Model 4  &  thin + thick & 1.06 & 1.59 & 0.98 & 11.88 & 2.76 \\
9 & yes & Model 4  &  thin + thick & 1.20 & 1.15 & 1.48 & 5.58 & 2.03  \\
10 & yes & Model 4  &  thin + thick & 1.10 & 1.54 & 3.87 & 2.10 & 3.29  \\
11 & yes & Model 4  &  thin + thick & 1.22 & 1.49 & 6.07 & 3.97 & 5.17  \\
12 & yes & Model 4  &  thin + thick & 1.24 & 1.37 & 6.21 & 14.95 & 6.94
 \\

\noalign{\smallskip}
\hline\
\end{tabular}
}
\label{tab:otrosmodels}
\end{center}
\end{table*}

In Table  \ref{tab:otrosmodels} we provide the values for the different reduced  $\chi^{2}$-test performed.  The goodness of the fitting for Model 2 as above commented, in particular for a disk age of $9\,$Gyr,  is quantified by the close to 1 reduced $\chi^{2}$-test values obtained. A value of $\chi^{2}_{\rm r,entire}=1.12$ for the overall luminosity function is achieved, while also close to 1 values are obtained for each of its different parts.

 Thus, we concluded that when a realistic distribution of metallicities spreading their values around the solar one is taken into account, detailed models including the sedimentation of $^{22}$Ne as well as thick disk contribution reproduce fairly well the different characteristics of the white dwarf luminosity function for an age of the thin disk around 9 Gyr.

\subsection{The effects of the age-metallicity relation}
\label{sec:casa}

In this section we analyze two additional models based on two different observed age-metallicity relations, namely: our Model 3, based on the observed age-metallicity relation presented by \citet{Casagrande2011,Casagrande2016,Haywood2013,Bergemann2014} and our Model 4, based on the age-metallicity relation provided by \cite{Twarog1980}. As previously stated on Section \ref{sec:metmod}, our  Model 3 consists of a linearly increasing metallicity for stars older than 8 Gyr, and a constant metallicity (solar with a dispersion of 0.4\,dex) for younger stars. Additionally,  our Model 4 predicts a monotonous increase of [Fe/H] beginning with a zero value for the oldest stars and  finishing with a solar metallicity value for present day stars.

\begin{figure}
   \resizebox{\hsize}{!}
   {\includegraphics[width=\columnwidth]{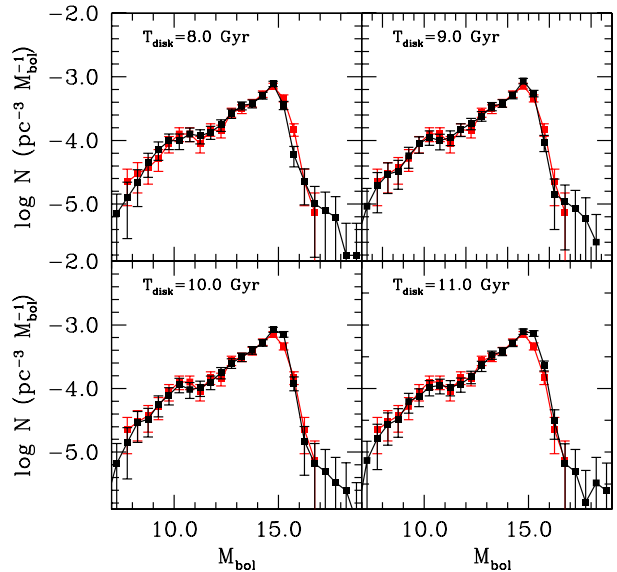}}
\caption{Same as Fig.\ref{f:model1} but considering our Model 3, based on \cite{Casagrande2011}.}
\label{f:model3}
\end{figure}

In Figure \ref{f:model3} we display the synthetic luminosity function resulting from our Model 3 (black line and dots) compared with the observed luminosity function (red line and dots)  of \cite{Limoges2015}. Additionally, the corresponding $\chi^2$ values are presented in Table \ref{tab:otrosmodels}. As can be seen from Fig.\ref{f:model3}, the synthetic luminosity function reproduces the observed features with high accuracy. That is indeed shown by the close to 1 reduced $\chi^2$ values achieved by the fits for the different regions considered, specifically for an age of 9 Gyr we obtain a $\chi^{2}_{\rm r,entire}=1.30$ for the entire luminosity function.

\begin{figure}
   \resizebox{\hsize}{!}
   {\includegraphics[width=\columnwidth]{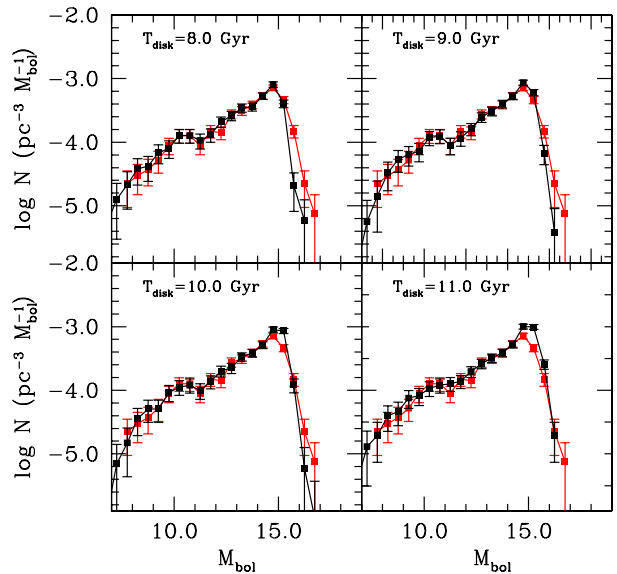}}
\caption{Same as Fig.\ref{f:model1} but considering our Model 4, based on Twarog's law.}
\label{f:model4}
\end{figure}

As a final exercise, our Model 4 reproduces the classical Twarog's age-metallicity model.  The synthetic and observed luminosity functions are shown in Figure \ref{f:model4}, while the corresponding $\chi^2$ tests are  provided in Table \ref{tab:otrosmodels}. A first look at Fig.\ref{f:model4} reveals that the general good agreement obtained with previous models (Models 2 and 3) is now lost. The best fit of the peak region is achieved for a disk age of $8\,$Gyr, while the cut-off is better fitted for an age around $10\,$Gyr. This clearly indicates the impossibility to find a good fit for any age for this model. Thus, we can conclude that an increasing age-metallicity model seems to hardly explain the observed characteristics of the local white dwarf luminosity function.


\subsection{The best-fit models}
\label{sec:bestfit}

\begin{figure}
   \resizebox{\hsize}{!}
   {\includegraphics[width=\columnwidth,clip=true,trim=0 70 0 100]{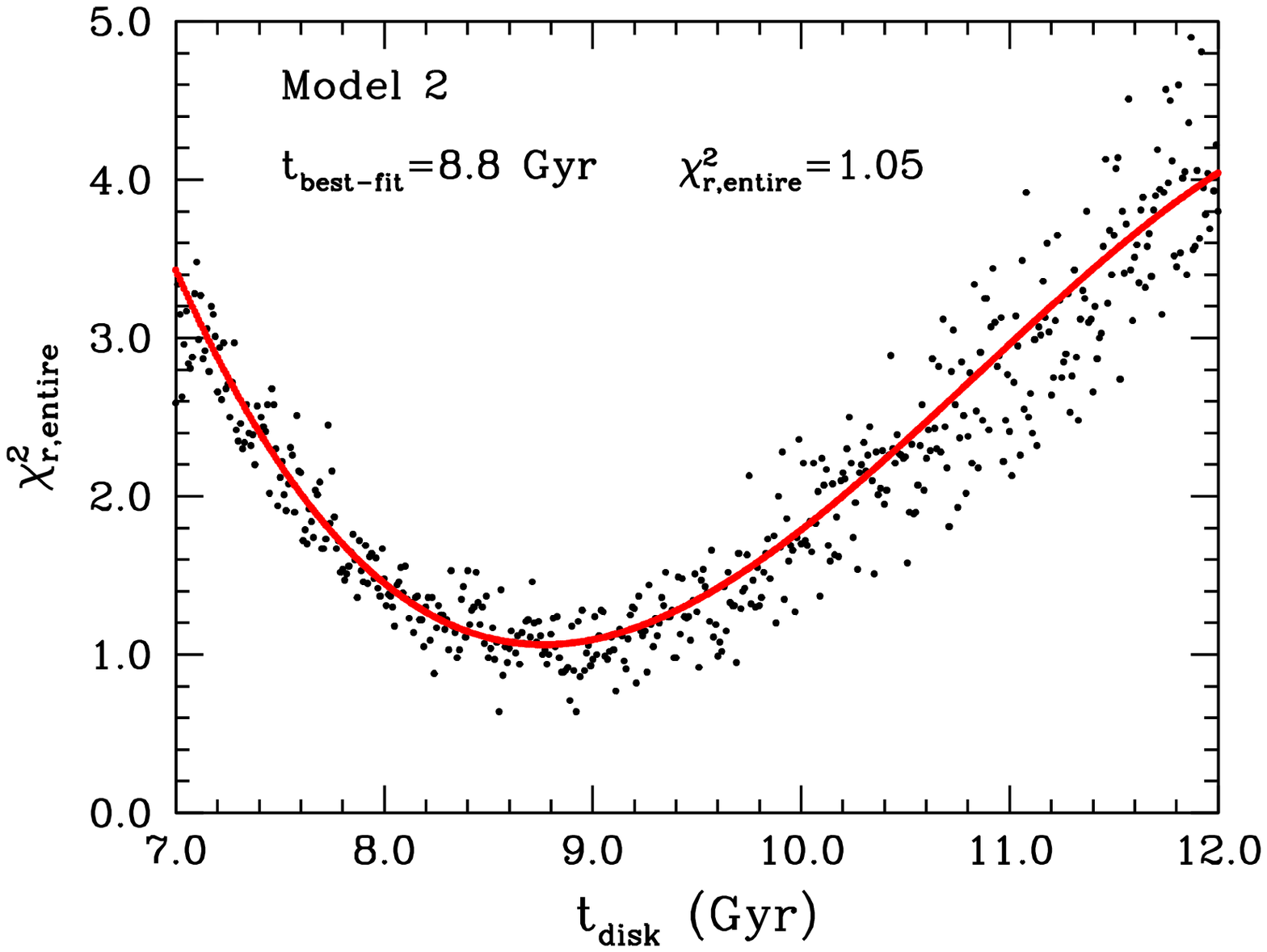}}
   {\includegraphics[width=\columnwidth,clip=true,trim=0 70 0 100]{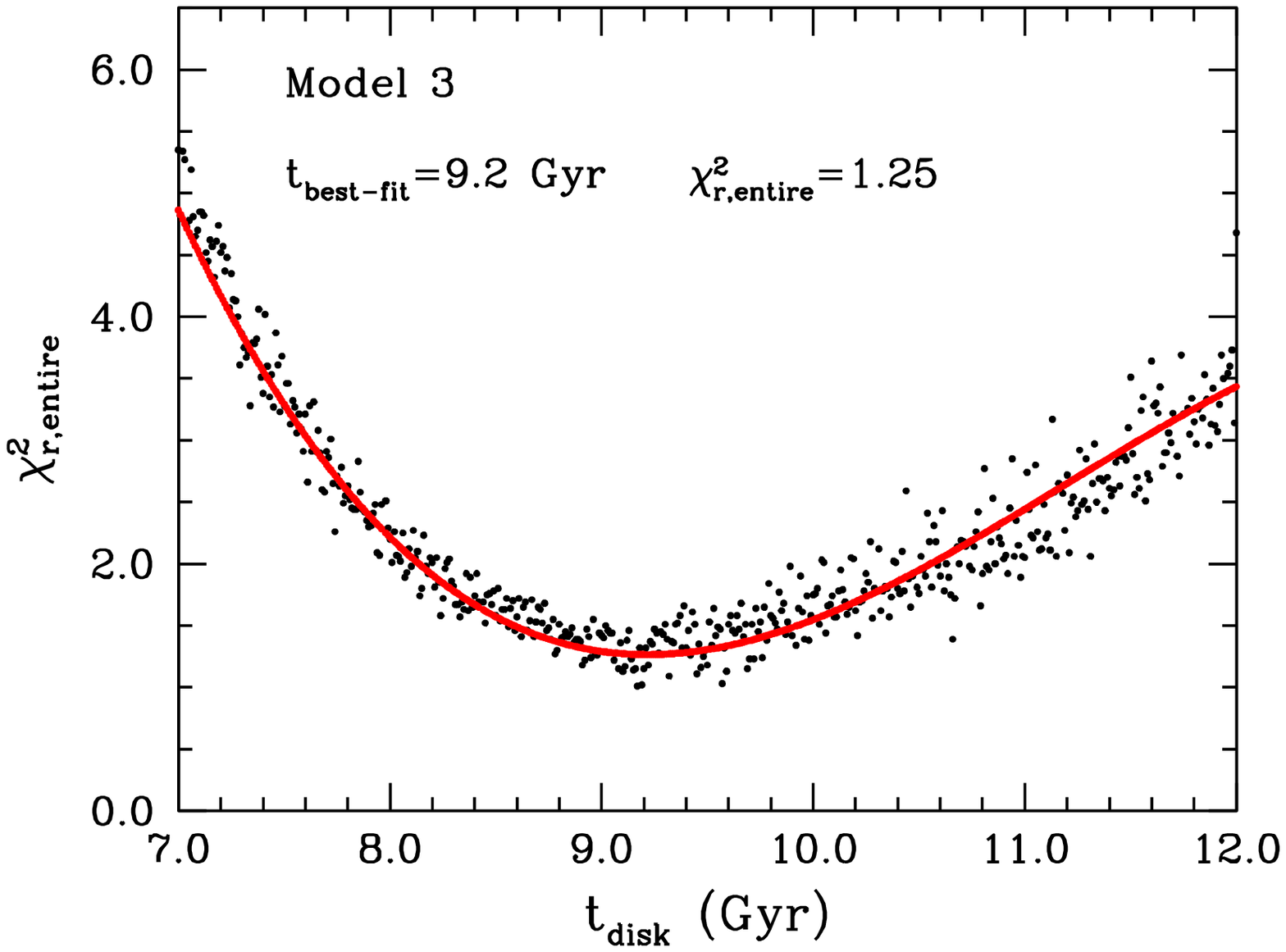}}
\caption{Reduced $\chi^2$ test value for the entire luminosity function as function of the disk age for Model 3 and Model 4. A polynomial fitting is represented as a red line. }
\label{f:chiages}
\end{figure}

Among the models studied so far, the best fits to the observational data are obtained for our Model 2 and Model 3-- these are the one that takes into account a metallicity dispersion around the solar value and the one  following \cite{Casagrande2011} data, respectively. We recall that both models include $^{22}$Ne sedimentation and a thick disk population. Given that both models reproduce with a high degree of accuracy the different regions of the white dwarf luminosity function, we aim to determine the disk age that best-fit the entire luminosity function. Thus, we ran our code for a wide range of ages between $7$ to $12\,$Gyr and by using a polynomial fit we find the minimum reduced $\chi^2$-test value. The resulting distributions for Models 2 and 3 are presented in Figure \ref{f:chiages} along with their corresponding polynomial fit (red line). By finding the minimum for these polynomials we derive a disk age of $T_{\rm disk}=8.8\pm0.2\,$Gyr for Model 2 and slightly larger, $T_{\rm disk}=9.2\pm0.3\,$Gyr  for Model 3. It is also worth saying that the best fit from Model 2 achieves a better statistical solution than the equivalent for Model 3:  $\chi^2_{\rm r,entire}$=1.05 for Model 2, while  $\chi^2_{\rm r,entire}$=1.25 for Model 3. 

\begin{figure}[t]
   {\includegraphics[width=1.05\columnwidth]{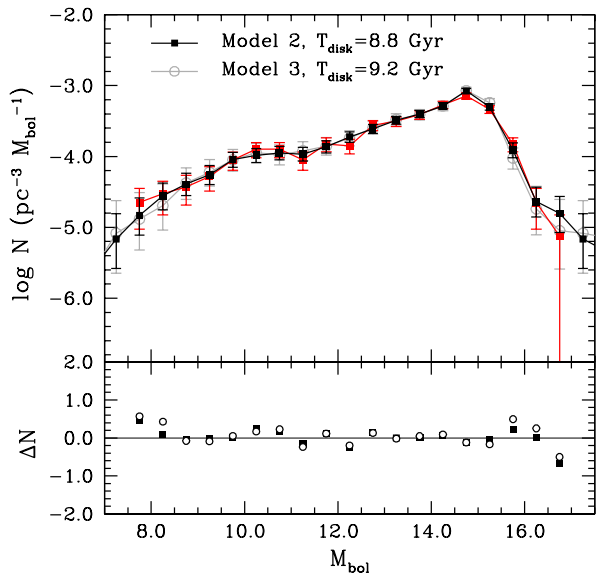}}
\caption{Synthetic white dwarf luminosity functions that result from our best-fit  metallicity  Model 2 (black solid squares) and Model 3 (gray  open circles), respectively, compared with the observed white dwarf luminosity function (red lines) of \cite{Limoges2015}. In the bottom panel we show the residuals between the simulated and the observed samples. See text for details.}
\label{f:bestfit}
\end{figure}

\begin{figure}
   \resizebox{\hsize}{!}
   {\includegraphics[width=1.05\columnwidth]{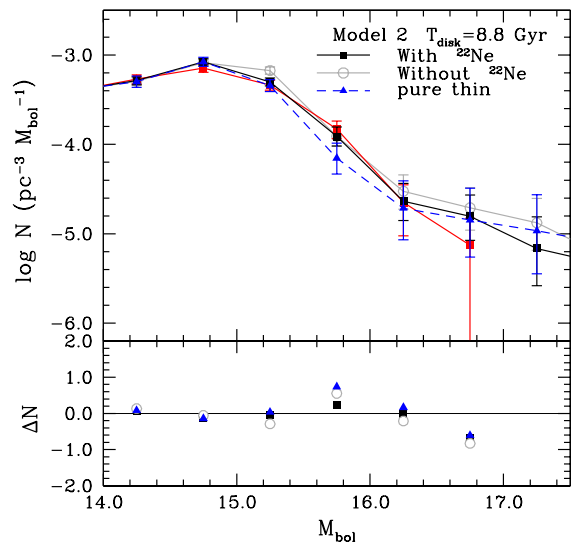}}
\caption{Faint end of the observed white dwarf luminosity function from \cite{Limoges2015} (solid square and continuous red line) compare to our best fit Model 2 when $^{22}$Ne sedimentation and a thick disk population is taken into account (solid square and continuous black line), when  $^{22}$Ne sedimentation is disregarded (open circle and continuous gray line) and when only pure thin disk population is considered (solid triangles and dashed blue line) }
\label{f:cutoff}
\end{figure}

As can be seen in Fig. \ref{f:bestfit}, the overall shapes of both synthetic luminosity functions excellently reproduce the observed data from  \cite{Limoges2015} -- red line. The different regions of the observed luminosity function are perfectly matched by our models, being slight better fitted by Model 2, as put into manifest for the closer to zero values of their residuals. However,  from an strict statistical point of view these small discrepancies are not enough to discard Model 3. The larger disk age derived from Model 3 is attributed to the lack of high metallicity stars during the first Gyr of formation in Model 3. But, as previously stated, given that both models are consistent with the observed data is not possible to discern the existence or not of suprasolar metallicity stars at the beginning of the thin disk formation. 

Finally, in order to summarize the effects of the different assumptions in our models, we analyze in bigger detail the faint downturn region of the luminosity function. In Figure \ref{f:cutoff} we plot the peak and cut-off regions of the observed luminosity function from \cite{Limoges2015} --red solid  square and continuous line-- compared to our Model 2 when $^{22}$Ne sedimentation and a thick disk population are taken into account --black solid square and continuous line--, when a thick disk population is included but $^{22}$Ne sedimentation is ignored --gray open circle and continuous line-- and when  $^{22}$Ne sedimentation  is considered but only for a pure thin disk population --blue solid triangle and dashed line. It seems clear from Fig. \ref{f:cutoff} that the best match is achieved when $^{22}$Ne sedimentation and a thick disk population are considered in a dispersion metallicity model. If, for instance, the thick disk population is ignored, a substantial lack of objects appears in the bin interval centered at $M_{\rm bol}=15.75$. Consequently, this excess of observed objects with respect to a pure thin disk population is,  as previously analyzed in Section \ref{s:thick}, a  clear signature of an old burst population. On the other hand, if we include a thick disk population but disregard the $^{22}$Ne sedimentation, the resulting luminosity function does not properly recover the observed distribution.  In this situation, an excess of objects appears in the faintest bins and also in the interval centered at $M_{\rm bol}=15.25$. As prompted in Section \ref{sec:e22ne}, the lack of an extra source of energy, as it is provided by the diffusion of $^{22}$Ne, entails an increase of the number of objects in the faintest bins. The quantitative analysis of the $\chi^2$ test shows a reasonably acceptable value for the entire luminosity function,  $\chi^2_{\rm r,entire}$=1.85, but a less good fitting in the peak region, $\chi^2_{\rm r,peak}$=2.00. Definitely, although the effects  of a thick disk population and the sedimentation of  $^{22}$Ne are of minor order, their inclusion in the modeling permits a better fitting of the observed data.

\section{Conclusions}
\label{sec:con}

Using  an state-of-the-art  Monte Carlo  simulator which incorporates the most up-to-date white dwarf evolutionary sequences, we have studied possible effects of cooling time delays induced by $^{22}$Ne sedimentation in the local white dwarf luminosity function along with different age-metallicity relation models and the possible contamination of a thick disk population. First of all, we analyzed a hypothetical scenario in which all stars belong to a pure thin disk population and have the same metallicity, fixed to the solar value. In this case, the synthetic models that result from incorporating and excluding $^{22}$Ne sedimentation perfectly match the observational data in the hot and medium region of the luminosity function. In these regions the effects of $^{22}$Ne sedimentation are negligible.  A small discrepancy arises between models  at the peak and cut-off regions. Although neither of the two models is able to adequately recover the drop-off  of the observed luminosity function for any of the disk ages studied here, we find hints that the model including $^{22}$Ne sedimentation seems to fit better the peak region. This fact is understood as a consequence of the extra release of energy due to $^{22}$Ne sedimentation which implies that more objects pile up in brighter magnitude intervals.

When a thick disk population is included in our models, the resulting effect appears as an increase of the number of stars in the interval centered around $M_{\rm bol}=15.75$. The fact that thick disk stars are modeled according to a single old burst formation implies that these stars concentrate to a specific magnitude interval, achieving in some cases more than $60\%$ of the contribution in this interval. Consequently, the inclusion of a thick disk population does not allow to fully recover the observed distribution of objects in the entire cut-off region, but improves the fitting in this region.

We  considered additional metallicity models which incorporate a dispersion around our fixed solar metallicity value as well as different age-metallicity relations.  We found that models including a metallicity dispersion  perfectly match the observations and, in particular are able to excellently reproduce the cut-off region.  The combined effect of shorter life-time for high metallicity progenitors, besides that $^{22}$Ne sedimentation is also more efficient for suprasolar metallicities, are suggested as the main causes for the agreement with the observed data. On the other hand, our Model 4 -- which assumes Twarog's law of increasing metallicity with age -- may be ruled out due to its poor agreement in particular of the peak and cut-off region. Consequently, models that include a metallicity dispersion (Model 2 and Model 3), that incorporate $^{22}$Ne sedimentation and a thick disk population present the better statistical performance, excellently reproducing the faint end along with the rest of the luminosity function. This permits us to robustly constrain the age of the Galactic disk by fitting not only the cut-off but also the entire luminosity function.   In particular, our best-fit model assumes a dispersion around the solar metallicity value  and an age of $8.8\pm0.2$ Gyr for the thin disk. The derived age for the thick disk in our best fit model is  $10.4\pm0.2$ Gyr for a $1\,$Gyr single burst formation. Although the effects of $^{22}$Ne sedimentation and a thick disk population assumption in the best fit model are only minor, their inclusion  permit a better fitting of the peak and cut-off regions.   Additionally, our Model 3, that is modeled according to \cite{Casagrande2011} data, although obtaining a slight worse performance is not disposable. The derived disk age for this model becomes slightly older, $9.2\pm0.3$ Gyr.

The luminosity function arising from our best-fit models reproduces with high accuracy all observed features: the bump at ${\rm M_{\rm bol}}\approx 10.5$ (assuming a recent enhanced of star formation 0.6\,Gyr ago, as claimed in \cite{Torres2016} as one plausible hypothesis), as well as the peak and the cut-off regions. For these models, discrepancies between the case in which $^{22}$Ne sedimentation is considered or disregarded are only marginal, although the model that takes into account  $^{22}$Ne sedimentation obtains  better values in the statistical tests. In the same sense, a thick disk population appears to better reproduce the magnitude interval centered around $M_{\rm bol}=15.75$ but it is only secondary in the fitting of the entire cut-off region. Definitely, metallicity effects due to a dispersion around the solar metallicity value can be pointed out as the most important factor for a proper reproduction of the faint end along with the peak of the local white dwarf luminosity function.


\begin{acknowledgements}

This work  was partially supported by the MINECO grant AYA\-2017-86274-P and the Ram\'on y Cajal programme RYC-2016-20254, by  the AGAUR,  by AGENCIA through  the  Programa de  Modernizaci\'on Tecnol\'ogica BID  1728/OC-AR, and by PIP  112-200801-00940 grant from CONICET. We acknowledge our anonymous referee for his/her detailed and valuable report.

\end{acknowledgements}

\bibliographystyle{aa}
\bibliography{nelf}

\end{document}